\documentclass[11pt]{article}
\usepackage{latexsym,cite}
\usepackage{graphicx}
\usepackage{bm}
\usepackage{xcolor}
\usepackage{float}
\usepackage{epsfig}

\usepackage{lineno,xcolor}
\usepackage{hyperref}

\usepackage{amsmath}

\newcommand{\be}{\begin{equation}}
\newcommand{\ee}{\end{equation} }
\newcommand{\beqa}{\begin{eqnarray}}
\newcommand{\eeqa}{\end{eqnarray} }
\newcommand{\ba}{\begin{array}}
\newcommand{\ea}{\end{array}}
\newcommand\dis{\displaystyle}

\newcommand{\half}{{{\textstyle\frac{1}{2}}}}

\newcommand\rd{{\rm d}}

\newcommand\kB{k_{{\scriptscriptstyle{\rm B}}}}

\newcommand{\red}[1]{{\color{red} #1 \color{black}}}

\newcommand{\blue}[1]{{\color{blue} #1 \color{black}}}

\newcommand{\purple}[1]{\textcolor[rgb]{0.8,0,1}{#1}}

\newcommand{\orange}[1]{{\color{orange} #1 \color{black}}}

\newcommand{\green}[1]{\textcolor[rgb]{0.2,0.5,0.1}{#1}}

\begin{document}
\begin{titlepage}
\title{\vskip -110pt
\vskip 2cm
Isobaric Critical Exponents: \\Test of  Analyticity  against NIST Reference  Data\\}
\author{\sc      Wonyoung Cho,  ~Do-Hyun Kim${}^{\ast}$ ~and~ Jeong-Hyuck Park${}^{\ast}$}
\date{}
\maketitle \vspace{-1.0cm}
\begin{center}
~~~\\
Department of Physics,
 Sogang University\\  35 Baekbeom-ro, Mapo-gu,
 Seoul 04107,   Korea\\
~~~\\
~~~\\
\end{center}
\begin{abstract}
\vskip0.1cm
\noindent
 Finite systems may undergo  first or second order phase transitions under not isovolumetric  but   isobaric  condition. The  `analyticity' of a finite-system partition function  has been argued to   imply
universal values for
\textit{isobaric critical exponents}, $\alpha_{{\scriptscriptstyle{P}}}$, $\beta_{{\scriptscriptstyle{P}}}$ and $\gamma_{{\scriptscriptstyle{P}}}$.
Here we test this prediction   by analyzing   NIST REFPROP data  for twenty major molecules, including $\mathrm{H_{2}O, CO_{2}, O_{2}}$,  \textit{etc.}  We  report  they are consistent  with the    prediction for  temperature range,   $10^{-5} <|T/T_{c}-1|<10^{-3}$.  For each molecule, there appears to exist  a characteristic natural number, $n=2,3,4,5,6$, which determines  all    the  critical exponents for $T<T_{c}$ as $\alpha_{{\scriptscriptstyle{P}}}=\gamma_{{\scriptscriptstyle{P}}}=\frac{n}{n+1}$ and $\beta_{{\scriptscriptstyle{P}}}=\delta^{-1}=\frac{1}{n+1}$.  For the opposite $T>T_{c}$,   all the fluids  seem to indicate   the universal value of ${n=2}$.
\end{abstract}

{\small
\begin{flushleft}
~~~~~~~${}^{\ast}$\textit{Co-correspondences}: dohyunkim@sogang.ac.kr \,\&\,   park@sogang.ac.kr\\
\end{flushleft}}
\thispagestyle{empty}
\end{titlepage}
\newpage

\section{Introduction}
\noindent The  celebrated paradigm  for  critical phenomena  in modern statistical physics  assumes
the thermo-dynamic limit~\cite{Anderson} with the intention of  achieving non-analyticity~\cite{YangLee}.
In this way, correlation length diverges at the critical point, which gives rise to the scaling theory of criticality. In order to describe the scaling, renormalization group methods were developed as for a mathematical framework, \textit{e.g.~}\cite{Fisher67, Heller67, Kadanoff67, Stanley, Wilson, Huang, Goldenfeld, Kadanoff00, Kadanoff09}.

However, the assertion  that the  thermodynamic limit for non-analyticity is a necessary condition to realize  phase transitions is valid only in the isovolumetric  case and
does not necessarily  apply to the isobaric situation~\cite{London,Imry,ChabaPathria,Wagner,Gross:1999jm,Gross:2002,Gross:2005sk}. In particular,  a  counterexample
based on the `analyticity' of a finite-system partition function has been suggested~\cite{7616relativistic}, which   
 predicts its own universal values for
isobaric critical exponents   defined in an analogous manner   to the usual ones except that the critical point is approached \textit{via}  isobaric transformations.

In order to boil water while heating, one may better
keep the pressure constant,  rather than fix the density  and take the thermodynamic limit.  In the low energy regime, the particle number, $N$, is conserved, and thus fixing the density is equivalent to keeping the volume constant.  In statistical physics, the partition function, $Z_{N}(T,V)$,  is  the sum of   Boltzmann factors, 
where the energy eigenvalues are generically quantized with respect to   the volume of the  system. One crucial difference  between  volume, $V$, and  pressure, $P$, is then that the former is a \textit{fundamental variable} of the partition function along with temperature, $T$, while the latter is a \textit{conjugate variable} derivable  from the partition function,
which is  \textit{a priori} a single-valued function of the two continuous  fundamental variables: $P(T,V)$.

For finite $N<<\infty$, the partition function is analytic, and hence, changing the temperature alone leaving the volume fixed cannot generate any singularity nor a discrete phase transition. To do so, the conventional  statistical physics  paradigm  suggests to  take the thermodynamic limit, $V\rightarrow\infty$ with the density, $N/V$, fixed. Yet, an alternative scheme has been also  around  in the literature, prescribing to keep \textit{not} the density \textit{but} the pressure constant, notably by London already in  1954~\cite{London}: For related discussions see  Imry \textit{et al.}~\cite{Imry}, Chaba \& Pathria~\cite{ChabaPathria} (theoretical),   Span \& Wagner~\cite{Wagner} (experimental), and Gross \textit{et al.}~\cite{Gross:1999jm,Gross:2002,Gross:2005sk} (micro-canonical ensemble). The pressure, not as the fundamental but as the conjugate variable, is to be fixed while the system undergoes any discrete phase transition.

In fact, under constant pressure condition, the analytic partition function of a finite system can feature a discrete phase transition provided the volume becomes a multi-valued ``function'' of $T$ and $P$. This happens when the partition function features a spinodal curve, which is defined
to be the collection of all the points on the $(T,V)$ plane where the isothermal compressibility becomes infinite, or
\be
\small{
\Phi(T,V):=\,-\,\frac{\partial P(T,V)}{\partial V}=0\,.}
\label{spinodal}
\ee
Equivalently, the second volume derivative of the free energy vanishes on the spinodal curve (\textit{e.g.}~\cite{Gibbs,Skripov,Chandler,Chimowitz,Connell,Shell}), which thus marks out the frontier of local stability.
Crucially, on the spinodal curve, the temperature derivative at  fixed pressure acting on an arbitrary physical quantity which is a function of $T$ and $V$, diverges since the isobaric  derivative contains the inverse, $\Phi^{-1}$,
\be
\ba{ll}
\dis{\left.\frac{\rd \,~}{\rd T}\right|_{P}}&\dis{ = \left.\frac{\partial~}{\partial T}\right|_{V} + \left.\frac{\rd V}{\rd T}\right|_{P} \left.\frac{\partial~}{\partial V}\right|_{T}}\\[12pt]
{}&\dis{=
\left.\frac{\partial~}{\partial T}\right|_{V}+\,\frac{~
\frac{\partial P(T,V)}{\partial T}
~}{\Phi(T,V)}
\left.\frac{\partial~}{\partial V}\right|_{T}
\,.}
\ea
\label{dTP}
\ee

The existence of the spinodal  curve  for a given analytic   partition function of a finite system has  been  verified: the standard textbook  systems of  ideal Bose gas  feature a spinodal curve, only if the number of the identical particles is sufficiently yet finitely large~~\cite{7616PRA,7616relativistic,Jeon:2011nk,Cho:2014joa,Park:2013gpa}.
Small perturbations due to extra interactions may well deform the shape of the spinodal curve but hardly eliminate its existence.   In the present paper,   we test the theoretical assertion  that  the spinodal curve exists,  or emerges~\cite{Park:2013gpa},  even for a generic interacting  system with finitely many particles, or  {real molecules},  while the corresponding partition function is analytic.
Under this assumption, it follows
that, \textit{the isothermal compressibility (or, equivalently, the inverse of $\Phi$)} is the only source of 
divergence when the isobaric derivative~(\ref{dTP}) acts on a  physical quantity which   is an analytic function of $T$ and $V$.  
This essentially  implies that the corresponding exponent of the singularity in such an isobaric situation is determined  by the power series expansion of $\Phi$ in a simple manner:
 the isobaric scaling behavior near a critical point, $(T_{c},V_{c})$, can be classified by  a natural number which   fixes  the  isobaric critical exponents in a universal and analytic manner, as we review below.

Specifically below,  we analyze   the (commercially available)  experimental data from ``NIST Reference Fluid Thermodynamic and Transport Properties Database (REFPROP)"~\cite{NIST}, which has been regarded as the world-best dataset for various fluids (especially $\mathrm{H_{2}O}$ and $\mathrm{CO_{2}}$ \cite{private}).
Among numerous molecules, we focus on twenty simple  molecules including $\mathrm{H_{2}O, CO_{2}, O_{2}}$,  \textit{etc.} and report that their critical exponents seem to  agree with the theoretical  prediction based on analyticity:  for each molecule, there is a characteristic natural number, $n=2,3,4,5,6$, which determines all the critical exponents for $T<T_{c}$ as $\alpha_{{\scriptscriptstyle{P}}}=\gamma_{{\scriptscriptstyle{P}}}=\frac{n}{n+1}$ and $\beta_{{\scriptscriptstyle{P}}}=\delta^{-1}=\frac{1}{n+1}$.  For the opposite $T>T_{c}$,   all the fluids  feature  the universal value of ${n=2}$.

\section{Review: Prediction from Analyticity}
\noindent We start with the  canonical partition function, $Z_{N}(T,V)$, of a generic $N$-particle system. For finite $N<<\infty$, the partition function is an \textit{analytic} function of the two continuous variables, $T$ and $V$.  The temperature dependence originates from the Boltzmann factor, $e^{-\beta E}$ with $\beta=(\kB T)^{-1}$, and the volume dependence arises due to the fact that the
energy eigenvalues are quantized with respect to   the size of the  system. Naturally,  all the physical quantities computable from the partition function are  \textit{a priori} functions of $T$ and $V$ too. For example, the  pressure is given by
\be
P(T,V)=\kB T\frac{\partial \ln Z_{N}(T,V)}{\partial V}\,.
\ee
It follows  infinitesimally,
\be
\rd P=\dis{\frac{\partial P(T,V)}{\partial T}\,\rd T+
\frac{\partial P(T,V)}{\partial V}\,\rd V}\,,
\label{dP}
\ee
and hence with the pressure kept fixed, \textit{i.e.~}on an isobar, we get
\be
\left.\frac{\rd T}{\rd V}\right|_{P}=-\left(\frac{\partial P(T,V)}{\partial V}\right)\left(\frac{\partial P(T,V)}{\partial T}\right)^{-1}\,.
\label{dTdV}
\ee

The spinodal curve (\ref{spinodal}) is defined to be the collection of all the points on the $(T,V)$ plane where the partial volume derivative of the pressure vanishes.
Physically, it corresponds to the boundary between the thermodynamically stable points, $\Phi>0$, and the unstable points, $\Phi<0$.  The existence of a spinodal  curve may not be  always guaranteed for a given (analytic)  partition function. Yet, it has  been shown,  first   numerically~\cite{7616PRA,7616relativistic} and  then through analytic   method~\cite{Jeon:2011nk,Cho:2014joa} that,  standard  textbook  systems of ideal Bose gas  feature a spinodal curve, and hence  the isobar on the $(T,V)$ plane zigzags,  if the number of particles is sufficiently large:  specifically $N\geq {7616}$ for the  $3D$ cubic box or $N\geq {35131}$ for the $2D$ square box.
Surely, these dimensionless critical numbers of particles are mathematically rather unnatural: the natural one would have been  $\infty$,~\textit{i.e.~}the thermodynamic limit.  But, the numbers are  quantum mechanically  generated    characteristics of  the geometric shape of the box, \textit{i.e.~}`cube' or `square'. Further,  their existence is anyhow physically natural, as  no experiment deals a  genuinely  infinite system.
Extra perturbative interaction may well  change the dimensionless  critical numbers and the shapes  of the spinodal curves.  Yet, what matters is the appearance  of a spinodal curve for large enough $N$~\cite{hasty}.  Hereafter,  even for a generic interacting  real system with finitely many particles,    we shall  assume,   merely that the spinodal curve exists or emerges, and further  tactically that  the temperature derivative of the pressure does not vanish (at least) on the spinodal curve,
\be
\left.\frac{\partial P(T,V)}{\partial T}\right|_{\Phi=0}\neq0\,.
\label{tactical}
\ee
It follows   from (\ref{dTdV}) that  $\left.\frac{\rd T}{\rd V}\right|_{P}$ vanishes on the spinodal curve, such  that
the isothermal  line ($\rd T=0$,  $\rd V\neq 0$) is  tangent to the isobar ($\rd P=0$).  This is also
consistent with the  definition of the spinodal curve~(\ref{spinodal}).

Crucially, on the spinodal curve, the temperature derivative at  fixed pressure acting on an arbitrary physical quantity which is a function of temperature and volume diverges, since the derivative contains the inverse, $\Phi^{-1}$, as shown in Eq.(\ref{dTP}).
This realizes a discrete phase transition  for a finite system under constant pressure condition.  Further,
if the physical quantity on which the derivative~(\ref{dTP}) acts   is an analytic function of $T$ and $V$, the inverse, $\Phi^{-1}$, is the only source of a divergence: none  of the   partial derivatives, neither  $\left.\frac{\partial~}{\partial T}\right|_{V} $  nor  $\left.\frac{\partial~}{\partial V}\right|_{T} $, can produce any singularity.  This  implies that the corresponding exponent of the singularity is determined by the power series expansion of $\Phi$ on the isobar and therefore, the  exponents can be classified  to take universal and exact
values,  under the assumptions of analytic partition function and isobaric transformations. 

The critical point, $(T_{c}, V_{c})$,  is  defined to be a particular point on the spinodal curve on which the higher order volume derivative of the pressure vanishes:
\be
\ba{l}
\dis{\Phi(T_{c},V_{c})=-\frac{\partial P(T_{c},V_{c})}{\partial V}}=0\,,\\[12pt]
\dis{\frac{\partial^{2}P(T_{c},V_{c})}{\partial V^{2}}=-\frac{\partial\Phi(T_{c},V_{c})}{\partial V}=0\,.}
\ea
\label{criticalprop1}
\ee
The first condition is the definition of the spinodal curve such that, as explained above,  the isothermal line  is tangent to the isobar, while the second condition  implies in a similar   fashion  that the isothermal line   is also  tangent to the spinodal curve. Thus,  the three curves, isothermal, spinodal and isobar, are all tangent to each other at a critical point. In general, for a certain natural number, $n\geq 2$, we let
\be
\ba{lll}
\dis{\frac{\partial^{k} P(T_{c},V_{c})}{\partial V^{k}}}=0\quad&\quad\mbox{for}\quad&\quad 1\leq k\leq n\,,
\ea
\label{criticalDEF1}
\ee
and
\be
\dis{\frac{\partial^{n} \Phi(T_{c},V_{c})}{\partial V^{n}}=-\frac{\partial^{n+1} P(T_{c},V_{c})}{\partial V^{n+1}}}\neq 0\,.
\label{criticalDEF2}
\ee

Now, at least locally,  around any spinodal point,  we   invert $P(T,V)$ to  express the temperature as a function of $P$ and $V$. This is justified by the assumption of \eqref{tactical} 
\textit{via} the implicit function theorem which also ensures that the function $T(P,V)$ is analytic as well.
We further  fix the pressure to be critical, \textit{i.e.~}$P_{c}=P(T_{c},V_{c})$.  This gives a local  expression of the temperature around the critical point  as a function of the volume under the fixed critical pressure,
\be
\ba{ll}
T=f_{c}(V)\,,\qquad&\qquad T_{c}=f_{c}(V_{c})\,,
\ea
\ee
satisfying
\be
P(f_{c}(V),V)=P_{c}\,.
\label{PfV}
\ee
The analyticity of the partition function, combined with the implicit function theorem, 
guarantees that $f_{c}(V)$ is 
an analytic function of $V$. By taking the volume-derivative of the relation~(\ref{PfV}), which should vanish, it is straightforward to show that the power series expansion of the function, $f_{c}(V)$,  around the critical volume starts at 
the $(n+1)$th order,
\be
T-T_{c}=\textstyle{\frac{1}{(n+1)!}}\left[\frac{\rd^{n+1}f_{c}(V_{c})}{\rd V^{n+1}}\right](V-V_{c})^{n+1}\,+\,\mbox{higher~orders}\,,
\label{Texp}
\ee
where, from our assumptions of (\ref{tactical}), (\ref{criticalDEF1}) and (\ref{criticalDEF2}), the leading order coefficient is nontrivial,
\be
\frac{\rd^{n+1}f_{c}(V_{c})}{\rd V^{n+1}}=\left[\frac{\partial^{n} \Phi(T_{c},V_{c})}{\partial V^{n}}\right]\left[\frac{\partial P(T_{c},V_{c})}{\partial T}\right]^{-1}\,\neq\,0\,.
\ee
Clearly, Eq.(\ref{Texp}) leads to the following isobaric   critical exponent:
\be
\ba{ll}
V/V_{c}-1~\sim~{\left|{T/T_{c}-1}\right|}^{\beta_{{\scriptscriptstyle{P}}}}\,,~~~&~~~\beta_{{\scriptscriptstyle{P}}}=\frac{1}{n+1}\,,
\ea
\label{beta}
\ee
which implies
\be
\left.\frac{\rd V}{\rd T}\right|_{P=P_{c}}\sim~{\left|{T/T_{c}-1}\right|}^{-\frac{n}{n+1}}\,.
\label{dVdT}
\ee
Similarly, with  (\ref{criticalDEF2}) and (\ref{Texp}),    the expansion of $\Phi$ on the critical isobar starts from the $n\,$th order in $V-V_{c}$ such that, in agreement with (\ref{dVdT}),
\be
\Phi(f(V),V)~~\sim~~(V/V_{c}-1)^{n}~~\sim~~{\left|{T/T_{c}-1}\right|}^{\frac{n}{n+1}}\,.
\label{divergence}
\ee
Thus, from (\ref{dTP}), \textit{any physical quantity given by the temperature derivative along the critical isobar  diverges  with the  universal  exponent,  $\frac{n}{n+1}$}. This includes the critical exponent of the specific heat  under constant pressure:
\be
\ba{ll}
C_{P}~\sim~{\left|{T/T_{c}-1}\right|}^{-\alpha_{{\scriptscriptstyle{P}}}}\,,~~~&~~~\alpha_{{\scriptscriptstyle{P}}}=\frac{n}{n+1}\,.
\ea
\label{alphaP}
\ee
On the other hand,  since  the specific heat  at constant volume or constant density is finite for any finite system,  the corresponding critical exponent should be trivial (if sufficiently zoomed in)~\cite{Wagner,zoomout}:
\be
\ba{lll}
C_{V}\,=\,\mbox{finite}~~&~~\mbox{\textit{i.e.}}~~&~~\alpha_{{\scriptscriptstyle{V}}}=0\,.
\ea
\label{alphaV}
\ee
Further,   the inverse of $\Phi$ gives the critical exponent of the isothermal compressibility on the critical isobar:
\be
\ba{ll}
\kappa_{{\scriptscriptstyle{T}}}:=-\frac{\partial\ln V(P_{c},T)}{\partial P}~\sim~{\left|{T/T_{c}-1}\right|}^{-\gamma_{{\scriptscriptstyle{P}}}}\,,~~&~~\gamma_{{\scriptscriptstyle{P}}}=\frac{n}{n+1}\,.
\ea
\label{gamma}
\ee
This isothermal compressibility exponent,
$\gamma_{{\scriptscriptstyle{P}}}=\frac{n}{n+1}$, is strictly  \textit{less} than one~(\ref{gamma}), and hence   differs from  the typical  estimations of its constant volume analog $\gamma_{\scriptscriptstyle{V}}$~\cite{Fisher67, Heller67, Kadanoff67, Stanley, Wilson, Huang, Goldenfeld, Kadanoff00}.
Nevertheless, the equality of  ${\alpha_{{\scriptscriptstyle{P}}}=\gamma_{{\scriptscriptstyle{P}}}}$ was   reported experimentally for some liquid-crystalline materials, such as   $n\mathrm{CB}$ and $\bar{m}\mathrm{S5}$~\cite{Wieczorek}.

Finally, from (\ref{criticalDEF2}), we  obtain the  \textit{isothermal} critical exponent at the critical temperature, ${T=T_{c}}\,$:
\be
\ba{ll}
P/P_{c}-1~\sim~{\left|{V/V_{c}-1}\right|}^{\delta}\,,~~~&~~~\delta=n+1\,.
\ea
\label{delta}
\ee
Note that, unlike  $\alpha_{{\scriptscriptstyle{P}}},\beta_{{\scriptscriptstyle{P}}}$ and $\gamma_{{\scriptscriptstyle{P}}}$, our last exponent, $\delta$,  is \textit{not} defined under the isobaric condition, and yet there should be no ambiguity in the definition~(\ref{delta}).

In summary,      under the combined hypotheses  of analyticity and existence of a spinodal curve,   we obtain  the following universal values for the (isobaric) critical exponents~\cite{7616relativistic}:
\be
\ba{ll}
\alpha_{{\scriptscriptstyle{P}}}=\gamma_{{\scriptscriptstyle{P}}}=\textstyle{\frac{n}{n+1}}\,,~~~&~~~~\beta_{{\scriptscriptstyle{P}}}=\delta^{-1}=\textstyle{\frac{1}{n+1}}\,.
\ea
\label{generalexponent}
\ee
They  are all determined by a single   natural number, $n$,  which should be greater than one, $n=2,3,4,5,\cdots$, and corresponds to the  characteristic  of a critical point.
Henceforth, we name the natural number, \textit{critical index}.
As can be easily checked, the isobaric critical exponents  satisfy the  `scaling laws'~\cite{Fisher67, Heller67, Kadanoff67, Stanley, Wilson, Huang, Goldenfeld, Kadanoff00}:
\be
\ba{rl}
\mbox{Rushbrooke}:~~&~~\alpha_{{\scriptscriptstyle{P}}}+2\beta_{{\scriptscriptstyle{P}}}+\gamma_{{\scriptscriptstyle{P}}}=2\,,\\
\mbox{Widom}:~~&~~\gamma_{{\scriptscriptstyle{P}}}=\beta_{{\scriptscriptstyle{P}}}(\delta-1)\,.
\ea
\ee

It is worthwhile to note that, at a generic spinodal point with  lower pressure, $P<P_{c}$, the case of  ${n=1}$   for (\ref{criticalDEF1}) and (\ref{criticalDEF2})  corresponds to the first-order phase transition between the liquid and the gas, which features the  superheating  and supercooling phenomena.  
Even at these points,  our analysis implies  the universal exponent:
\be
\alpha_{{\scriptscriptstyle{P}}}=\beta_{{\scriptscriptstyle{P}}}=\gamma_{{\scriptscriptstyle{P}}}=\delta^{-1}=\half\,.
\label{exponentnequalone}
\ee

We   stress that  our analysis and  the resulting  critical exponents~(\ref{generalexponent}) as well as (\ref{exponentnequalone}) should hold universally   not only for   liquid-vapor transitions  but also  for any isobaric phase transition  where its own  spinodal curve can be defined.   Such examples include  the Nematic to Smetic-A phase transition of the liquid-crystalline  experiment (without $\delta$)~\cite{Wieczorek}  and  a relativistic ideal Bose gas theory   (see Figure~5 of  \cite{7616relativistic}). Both examples  feature the same critical index, `$n=2$' and agree with (\ref{exponentnequalone}) when $P<P_{c}$.

\section{Van der Waals fluid: Toy Model of ${n}=2$}
\noindent The first historical example of mean field theory was the van der Waals fluid, which we show here   to provide a simple example of the above general consideration.
The van der Waals equation reads
\be
\left(P_{r}+\frac{3}{v_{r}^{2}}\right)\left(v_{r}-\frac{1}{3}\right)=\frac{8}{3}T_{r}\,,
\label{vdW}
\ee
where $P_{r}=P/P_{c}$, $v_{r}=v/ v_{c}$ and $T_{r}=T/T_{c}$ are reduced pressure, volume per particle and temperature, such that the critical point is given by  $P_{r}=v_{r}=T_{r}=1$.  Around the critical point, the critical isobar, $P_{r}\equiv 1$, is given by
\be
T_{r}=\frac{1}{8}(3v_{r}-1)\left(1+\frac{3}{v_{r}^{2}}\right)=1+\frac{3(v_{r}-1)^{3}}{8v_{r}^{2}}\,.
\label{vdWbeta}
\ee
Therefore, near the critical point,  $T_{r}-1\sim (v_{r}-1)^{3}$, we get  $n=2$, and hence $\beta_{{\scriptscriptstyle{P}}}=\frac{1}{3}$. On the other hand, taking the $v_{r}$-derivative  of (\ref{vdW}), with fixed temperature, gives
\be
\left(\frac{\partial P_{r}(v_{r},T_{r})}{\partial v_{r}}-\frac{6}{v_{r}^{3}}\right)\left(v_{r}-\frac{1}{3}\right)+P_{r}+\frac{3}{v_{r}^{2}}=0\,,
\ee
and hence, combined with \eqref{vdWbeta}, we get  $\gamma_{{\scriptscriptstyle{P}}}=\frac{2}{3}$, as follows
\be
\ba{ll}
\kappa_{{\scriptscriptstyle{T}}}&=-\left.\frac{\partial\ln v_{r}(P_{r},T_{r})}{\partial P_{r}}\right|_{P_{r}=1}=
\frac{v_{r}^{2}(v_{r}-\frac{1}{3})}{(v_{r}+2)(v_{r}-1)^{2}}\\
&
\sim~\frac{1}{(v_{r}-1)^{2}}~\sim~(T_{r}-1)^{-\frac{2}{3}}\,.
\ea
\ee
From (\ref{vdW}) the critical isothermal relation between pressure and  temperature reads, with $T_{r}\equiv1$,
\be
P_{r}-1=-\frac{(v_{r}-1)^{3}}{(v_{r}-\frac{1}{3})v_{r}^{2}}~\sim~(v_{r}-1)^{3}\,.
\ee
Thus, we note $\delta=3$. Lastly,  in order to obtain $\alpha_{{\scriptscriptstyle{P}}}$,
one should first remember that the specific heat at constant
volume of any fluid whose equation of state is linear in  temperature is equal to that of the classical ideal gas.
Thus it is finite and constant, so the corresponding critical exponent should be trivial, \textit{i.e.~}$\alpha_{{\scriptscriptstyle{V}}} = 0$~(\ref{alphaV}). On the other hand, the specific heat at constant pressure, $C_{P}$,  is given by the thermodynamic identity (see \textit{e.g.~}\cite{Goldenfeld}),
\be
C_{P} - C_{V} = - T_{r} \left( \left.\frac{\partial P_{r}}{\partial T_{r}} \right|_{v_{r}}\right)^{\!2} \left.\frac{\partial v_{r}}{\partial P_{r}} \right|_{T_{r}} ,
\ee
which is proportional to $\kappa_{{\scriptscriptstyle{T}}}$. Therefore, $\alpha_{{\scriptscriptstyle{P}}}$, the  critical
exponent of the specific heat under constant pressure takes  the same value as $\gamma_{{\scriptscriptstyle{P}}} = \frac{2}{3}$.     Put all together,    the critical point of  van der Waals fluid belongs to the class  of $n=2$.
Note that the  textbook
approach to obtain the  critical exponents of van der Waals fluid is based on the concept of Maxwell's equal area construction \cite{Huang,Goldenfeld, Sengers74},
which, we assert,  cannot be applied nor defined at the critical point itself. More precisely, the conventional derivation of the exponent, ``$\beta=\frac{1}{2}$" is based on the behavior of the \textit{binodal} curve of liquid-vapor coexistence near the critical point,
while our  isobaric critical exponent, $\beta_{{\scriptscriptstyle{P}}}=\frac{1}{3}$,  is  obtained here by simply fixing the pressure to be critical, $P_{r}\equiv 1$,  in the very van der Waals equation~(\ref{vdWbeta}). 

\section{Result and Discussion}
\noindent We  recall the theoretical prediction of the critical exponents~(\ref{generalexponent}), $\alpha_{{\scriptscriptstyle{P}}}=\gamma_{{\scriptscriptstyle{P}}}=\frac{n}{n+1}$, $\beta_{{\scriptscriptstyle{P}}}=\delta^{-1}=\frac{1}{n+1}$, which   can be easily derived   \textit{not} by assuming  the thermodynamic limit  nor any singularity of the free energy, \textit{but} by postulating   the analyticity of the partition function and the existence of a spinodal curve. Our main result   in this work is  to verify  that such a simple  approach appears  relevant  to the description of the  critical phenomena of the liquid-gas phase transition of `real fluids': the analyticity-based theoretical prediction is consistent with  the high-precision experimental data, REFPROP of  NIST~\cite{NIST}.  

The NIST  data are sufficiently precise even close to  the critical points: the uncertainties for most fluids do not exceed  2\%. 
For examples, the uncertainty in pressure of nitrogen in the critical region is estimated to be 0.02\%. The uncertainty in pressure of water in the critical region is 0.1\%. The uncertainty in heat capacities of argon is within 0.3\% for the vapor and 2\% for the liquid. The uncertainties in vapor pressure of helium are less than 0.02\% and for the heat capacities are about 2\%.  
The detailed information on the data can be obtained by downloading the (commercially available) REFPROP database program through  its website~\cite{NIST}.
We tend to believe that such high-precision  data were  not available while the conventional paradigm of taking the thermodynamic limit and achieving  non-analyticity  was proposed and established during the mid 20th century. In particular,  the isothermal relation between pressure and volume appears to have been  rather poorly measured, \textit{e.g.~}\cite{Heller67}.

\begin{figure}[H]
\includegraphics[width=140mm]{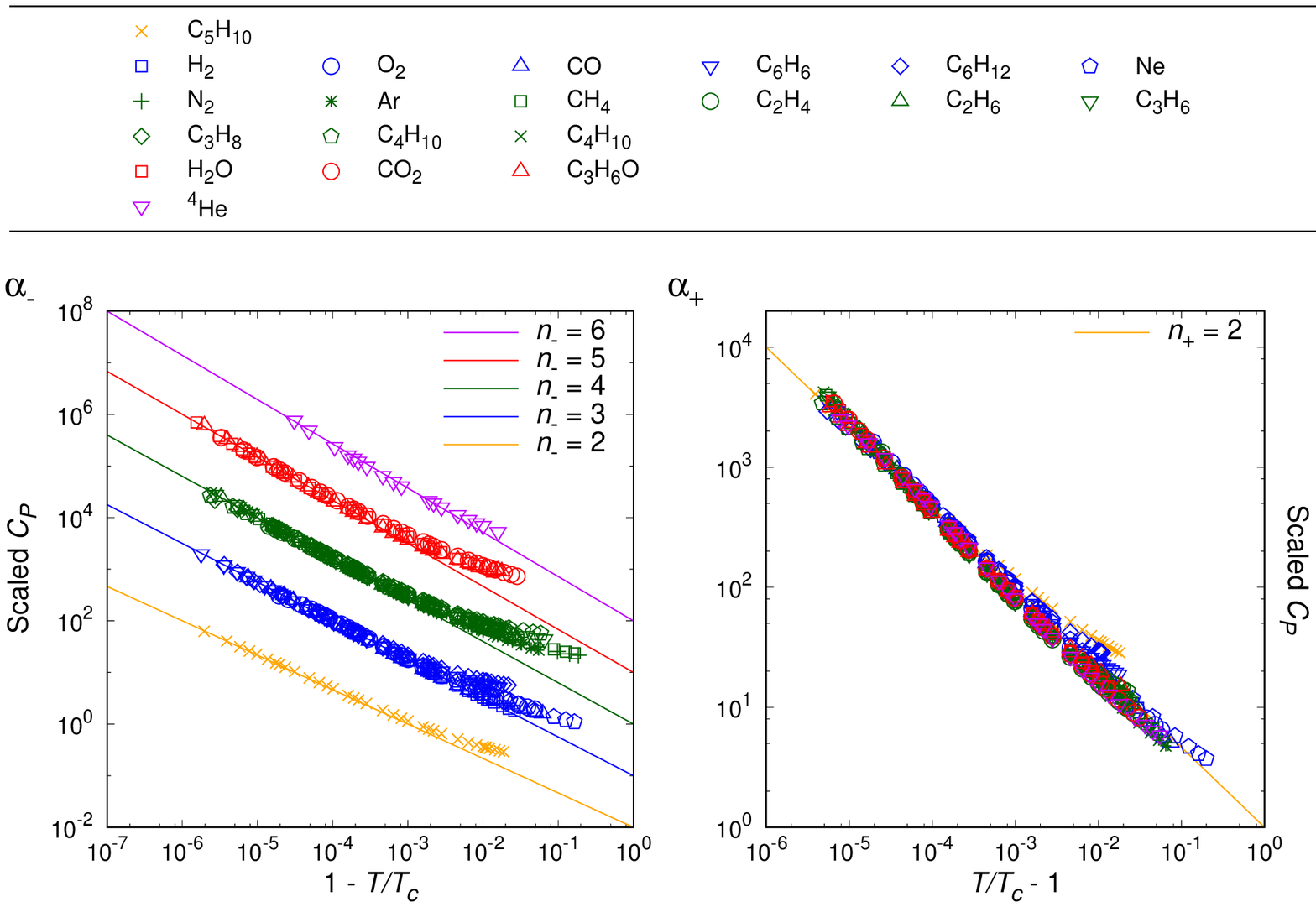}
\caption{{{{\sffamily Critical exponent, $\alpha_{{\scriptscriptstyle{P}}}=\frac{n}{n+1}$\,:}}} {{ \sffamily $n_{-}=2,3,4,5,6$ for $T<T_{c}\,$ and $\,n_{+}=2$ for $T>T_{c}\,$}}}
\label{FIGalpha}
\end{figure}

\begin{figure}[H]
\includegraphics[width=140mm]{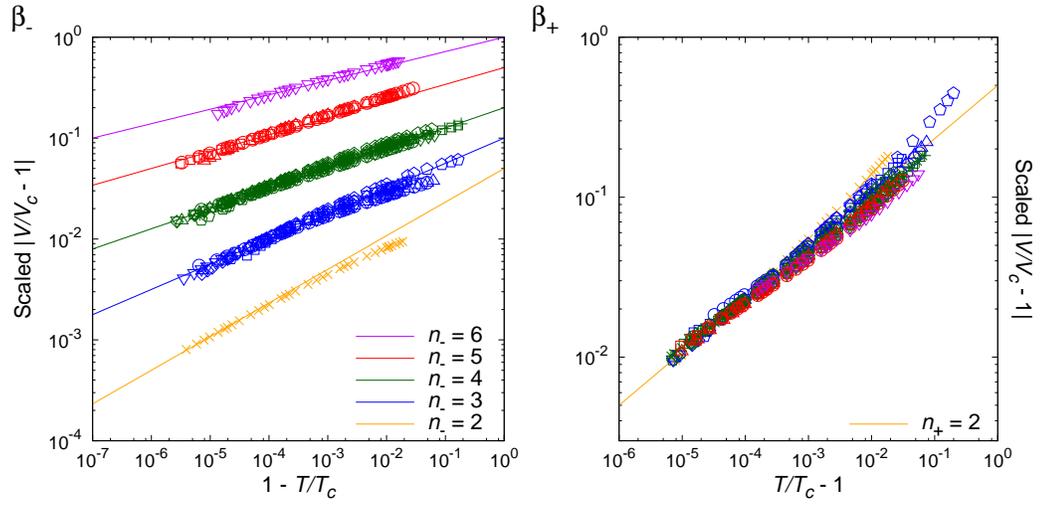}
\caption{{{{\sffamily Critical exponent, $\beta_{{\scriptscriptstyle{P}}}=\frac{1}{n+1}$\,:}}} {{\sffamily $n_{-}=2,3,4,5,6$ for $T<T_{c}\,$ and $\,n_{+}=2$ for $T>T_{c}\,$}}}
\label{FIGbeta}
\end{figure}

\begin{figure}[H]
\includegraphics[width=140mm]{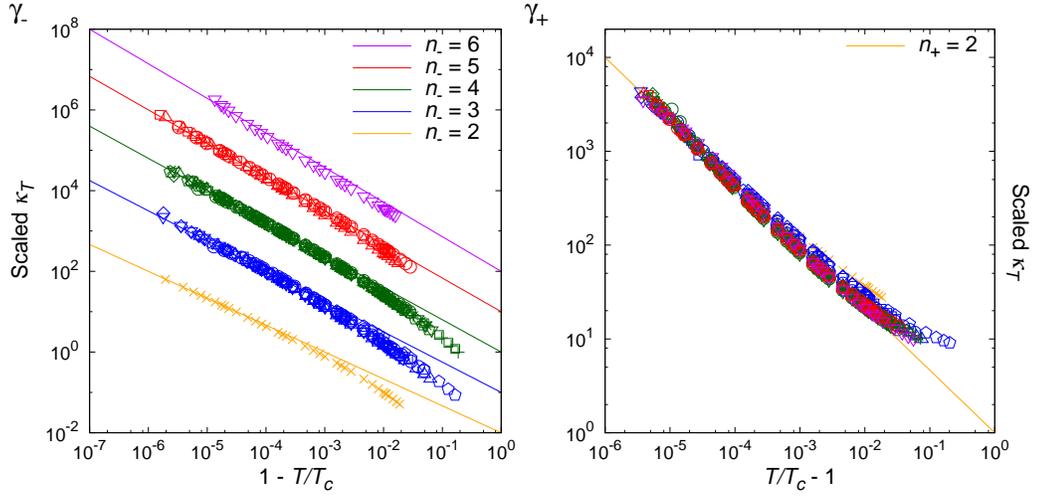}
\caption{{{{\sffamily Critical exponent, $\gamma_{{\scriptscriptstyle{P}}}=\frac{n}{n+1}$\,:}}} {{\sffamily $n_{-}=2,3,4,5,6$ for $T<T_{c}\,$ and $\,n_{+}=2$ for $T>T_{c}\,$}}}
\label{FIGgamma}
\end{figure}

\begin{figure}[H]
\includegraphics[width=140mm]{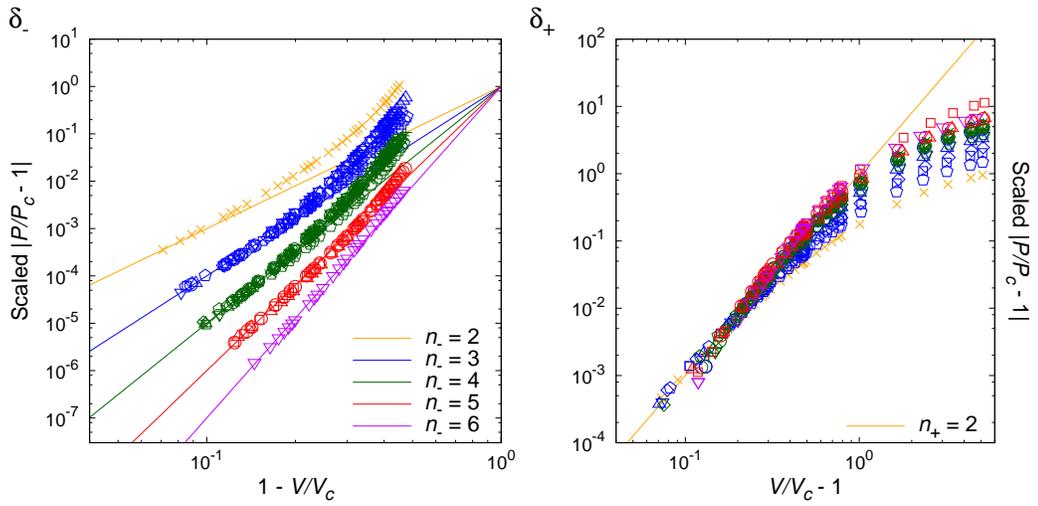}
\caption{{{{\sffamily Critical exponent, $\delta={n+1}$\,:}}} {{\sffamily $n_{-}=2,3,4,5,6$ for $V<V_{c}\,$ and $\,n_{+}=2$ for $V>V_{c}\,$}}}
\label{FIGdelta}
\end{figure}

By analyzing the data for  randomly  chosen twenty simple molecules,  we are able  to classify the critical phenomena of  those real fluids in terms of the critical index, $n_{-}=2,3,4,5,6$,  for $T<T_{c}$. For the opposite range of temperature,   $T>T_{c}$, the natural number seems to be universal as ${n_{+}=2}$, which agrees with van der Waals fluid~(\ref{vdW}), a relativistic ideal Bose gas ($n_{-}=n_{+}=2$~\cite{7616relativistic}), and an experimental result on $\mathrm{SF}_{6}$~\cite{Wagner2}.  
Table~\ref{TableCOMPARE} and Figures~\ref{FIGalpha},~\ref{FIGbeta},~\ref{FIGgamma},~\ref{FIGdelta} show the main results.

Table~\ref{TableCOMPARE} contains our  classification of the twenty molecular  fluids,  plus the relativistic ideal Bose gas and the van der Waals fluid, according to the critical indices: $n_{-}$ for $T<T_{c}$ and $n_{+}$ for $T>T_{c}$. Note the universal value,  ${n_{+}=2}$. We have determined the critical indices from the experimental data of NIST  (REFPROP)~\cite{NIST} especially over the reduced temperature range, $10^{-5}<\left|T/T_{c}-1\right|<10^{-2}$.  

Figures~\ref{FIGalpha},~\ref{FIGbeta},~\ref{FIGgamma} show the log-log plot of the scaled physical quantities ($C_{P}$, $|V/V_{c}-1|$, and $\kappa_{T}$) \textit{versus}  the reduced temperature ($| T/T_{c} -1 |$) for  the twenty  molecules, where data are logarithmically-binned from REFPROP dataset for our analysis. On the other hand, Figure~\ref{FIGdelta} shows the log-log plot of the scaled reduced pressure ($|P/P_{c}-1|$) \textit{versus} the reduced volume ($|V/V_{c}-1|$) for the same molecules, where data are also logarithmically-binned.
All the  figures represent the best fit plots of experimental data obtained from the twenty molecules, which confirm the scaling behaviour characterized  by the critical indices, $n_{-}=2,3,4,5,6$ for $T<T_{c}$ and ${n_{+}=2}$ for $T>T_{c}$.

Generically, except $\mathrm{C}_{5}\mathrm{H}_{10}$,     we have obtained  two distinct critical indices, $n_{-}$ and $n_{+}$,  depending on $T<T_{c}$ and  $T>T_{c}$.
This  rather surprising result leads us to speculate  ---still assuming the analyticiy--- that   there may  exist   more than one critical points quite  close to each other, such as $(T_{c-},V_{c-}:n_{-})$ and $(T_{c+},V_{c+}:n_{+})$~\cite{preparation}. 

We may regard our result as a tantalizing hint  for  the finiteness   of  the  Avogadro number in real  experimental laboratories, 
such that  analyticity appears to persist    at least  for the close up range like $ 10^{-5}<|T/T_{c}-1|< 10^{-3}$.  On the other hand, slightly further away from  the critical point, \textit{e.g.~}$ 10^{-3}<|T/T_{c}-1|< 10^{-1}$, the singularity or  the critical index appears to get \textit{enhanced} still in a universal manner for $\beta_{-}, \gamma_{-},\delta_{-}$. This    might correspond to a thermodynamic limiting behavior.


 As a matter of fact, NIST REFPROP assumes certain form of extrapolation  formulas for data fitting, such that it appears able to  generate data for  extremely  short range, \textit{e.g.~}$10^{-9}<|T/T_{c}-1|<
10^{-5}$.  Yet, such  data turn out to deviate from the analytic prediction of our interest, which  should be  due to the artificially designed fitting formulas. While for the range of $10^{-5}<|T/T_{c}-1|< 10^{-3}$, we trust  NIST  REFPROP represents  true experimental data~\cite{private}.  For this range   it is a highly nontrivial verification of us   that  \textit{all   the four   critical exponents of a given real  molecule are fixed  by a `common'  natural number, \textit{i.e.~}the critical index}.

\section{Conclusion}
Finite systems can undergo  first or second order phase transitions under isobaric  condition without resorting to the thermodynamic limit. 
The  `analyticity' of a finite-system partition function    predicts then the  universal values of the 
isobaric critical exponents. 
 In this work,  analyzing   NIST REFPROP data  for twenty simple molecules, including $\mathrm{H_{2}O, CO_{2}, O_{2}}$,  \textit{etc.} over the  the reduced temperature range of  $10^{-5} <|T/T_{c}-1|<10^{-3}$,   we have tested  and verified the prediction:  for each molecule, there appears to exist  a characteristic natural number, $n=2,3,4,5,6$, which determines  all    the  critical exponents for $T<T_{c}$ as $\alpha_{{\scriptscriptstyle{P}}}=\gamma_{{\scriptscriptstyle{P}}}=\frac{n}{n+1}$ and $\beta_{{\scriptscriptstyle{P}}}=\delta^{-1}=\frac{1}{n+1}$.
For the opposite $T>T_{c}$,   all the fluids  seem to indicate   the universal value of ${n=2}$.

 In the sense of Kuhn~\cite{Kuhn},  our result might be viewed as ``anomaly"  within  the standard paradigm of taking the thermodynamic limit  for critical phenomena,  \textit{e.g.~}the (non-analytic)  $3D$ Ising model or   $XY$ model \cite{Garland}.  Nevertheless, our analysis  represents truthfully the NIST experimental data for the specific  temperature window,  $10^{-5}<|T/T_{c}-1|< 10^{-3}$. We call for further theoretical as well as  experimental  investigations. 

\begin{table}[H]
\caption{Classification of the $20$ major fluids  plus the relativistic ideal Bose gas and the van der Waals fluid, according to the critical indices: $n_{-}$ for $T<T_{c}$ and $n_{+}$ for $T>T_{c}$.
}
\label{TableCOMPARE}
\begin{center}
\begin{tabular}{|c|c|}
\hline
{\bf{\begin{tabular}{c}\!Critical Index\!\\
$~\mathbf{(n_{-},n_{+})}~$\end{tabular}}}&~~{\bf{Fluids}}~~\\
\hline
$\orange{\mathbf{(2,2)}}$&\begin{tabular}{c}
\orange{$\mathrm{C}_{5}\mathrm{H}_{10}$ (cyclopentane),} \\
\orange{van der Waals fluid,} \\
\orange{relativistic ideal Bose gas~\cite{7616relativistic}}
\end{tabular}\\
\hline
$\blue{\mathbf{(3,2)}}$&
\blue{\begin{tabular}{ll}
$\mathrm{H}_{2}$ (hydrogen), & $\mathrm{O}_{2}$ (oxygen), \\
 $\mathrm{CO}$ (carbon monoxide), & $\mathrm{C}_{6}\mathrm{H}_{6}$ (benzene), \\
  $\mathrm{C}_{6}\mathrm{H}_{12}$ (cyclohexane), & $\mathrm{Ne}$ (neon)
 \end{tabular}}\\
 \hline
$\green{\mathbf{(4,2)}}$&
\green{\begin{tabular}{ll}
$\mathrm{N}_{2}$ (nitrogen), ~~~~~\quad~~~& $\mathrm{Ar}$ (argon), \\
 $\mathrm{CH}_{4}$ (methane), ~& $\mathrm{C}_{2}\mathrm{H}_{4}$ (ethylene), \\
  $\mathrm{C}_{2}\mathrm{H}_{6}$ (ethane), ~& $\mathrm{C}_{3}\mathrm{H}_{6}$ (propylene),\\
  $\mathrm{C}_{3}\mathrm{H}_{8}$ (propane), ~&
  $\mathrm{C}_{4}\mathrm{H}_{10}$ (butane),\\
  \multicolumn{2}{c}{$\mathrm{C}_{4}\mathrm{H}_{10}$ (isobutane)}
 \end{tabular}}\\
 \hline
$\red{\mathbf{(5,2)}}$&
\red{\begin{tabular}{ll}
$\mathrm{H}_{2}\mathrm{O}$ (water), ~~~& $\mathrm{CO}_{2}$ (carbon dioxide),\\
 \multicolumn{2}{c}{$\mathrm{C}_{3}\mathrm{H}_{6}\mathrm{O}$ (acetone)}
 \end{tabular}}\\
 \hline
$\purple{\mathbf{(6,2)}}$&~\purple{${}^{4}\mathrm{He}$ (helium-4)}\\
\hline
\end{tabular}
\end{center}
\vspace*{-12pt}
\end{table}


\vspace*{-20pt}

\vspace*{-5pt}

\newpage

\vspace*{-5pt}

\vspace*{-5pt}

\section*{Acknowledgements}

{We wish to thank Xavier Bekaert for numerous  enlightening discussions and  Eric W. Lemmon at NIST for helpful correspondences.} {This work was  supported by Sogang University (Grant Nos. 201610033.01 and 201710066.01) and the National Research Foundation of Korea (Grant Nos. 2015K1A3A1A21000302 and 2016R1D1A1B01015196).}\\

{JHP proposed the research.   WYC contributed  the  NIST data handling and analysis.   DHK led the interpretation of the data.  JHP and DHK wrote the manuscript.} {The authors declare no competing interests.}

\section{Appendix: Comment on Guggenheim's fitting}
\noindent Around the critical point,  with $\Phi_{c}\equiv\Phi(T_{c},V_{c})=0$ and from  (\ref{criticalDEF2}),    we  may expand
\be
\ba{ll}
\Phi(T,V)
=&{\left(\frac{\,\partial\Phi_{c}}{\partial T}
+\frac{\partial^{2}\Phi_{c}}{\partial T\partial V}{\Delta V}+\frac{1}{2}\frac{\partial^{2}\Phi_{c}}{\partial T^{2}}{\Delta T}\right){\Delta T}}\\
{}&{+\frac{1}{n!}\frac{\partial^{n} \Phi_{c}}{\partial V^{n}}{\Delta V}^{n}\,+\cdots\,,}
\ea
\ee
where  $\Delta T={T-T_{c}}$, $\,\Delta V={V-V_{c}}$.
Then,  near to the critical point, in comparison with the critical isobar~(\ref{beta}),  the  spinodal curve can be approximated by
\be
0~<~T/T_{c}-1~\sim~(V/V_{c}-1)^{{n}}\,.
\label{spinodalTV}
\ee
Our result  implies then that,  the supercooling ($T<T_{c}$) and the superheating ($T>T_{c})$ spinodal curves read separately   in terms of the density, $\rho_{\pm}=N/V_{\pm}$,
\be
\rho_{\pm}/\rho_{c}-1~\sim~|T/T_{c}-1|^{{1}/{n_{\pm}}}\,.
\ee
From the empirical result reported  in this work, $ n_{+}< n_{-}$
(except  $\mathrm{C}_{5}\mathrm{H}_{10}$),   we expect  close   to the critical point,
\be
|T/T_{c}-1|^{{1}/{n_{+}}} << |T/T_{c}-1|^{1/n_{-}}\,.
\ee
Thus,  the difference between the densities of the supercooled gas and the superheated liquid can be approximated by
\be
\rho_{+}-\rho_{-}~\sim~|T/T_{c}-1|^{{1}/{n_{-}}}\,.
\ee
This    is comparable with the    renowned  Guggenheim's fitting~\cite{Guggenheim45} with $n=3$.\\










\begin{thebibliography}{99}
\bibitem{zoomout}
On the other hand,   if appropriately  zoomed out, it might be possible to observe   pseudo-thermodynamic-limiting  behavior, with finite peak.
\end{thebibliography}
\end{document}